\documentstyle[12pt]{article}

\font\twlgot =eufm10 scaled \magstep1 \font\egtgot =eufm8
\font\sevgot =eufm7 \font\twlmsb =msbm10 scaled \magstep1
\font\egtmsb =msbm8 \font\sevmsb =msbm7

\newfam\gotfam
\def\pgot{\fam\gotfam\twlgot}
\textfont\gotfam\twlgot \scriptfont\gotfam\egtgot
\scriptscriptfont\gotfam\sevgot
\def\got{\protect\pgot}
\newfam\msbfam
\textfont\msbfam\twlmsb \scriptfont\msbfam\egtmsb
\scriptscriptfont\msbfam\sevmsb
\def\Bbb{\protect\pBbb}
\def\pBbb{\relax\ifmmode\expandafter\Bb\else\typeout{You cann't use
Bbb in text mode}\fi}
\def\Bb #1{{\fam\msbfam\relax#1}}

\def\thebibliography#1{\section*{References}\list
  {[\arabic{enumi}]}{\settowidth\labelwidth{#1}\leftmargin\labelwidth
    \advance\leftmargin\labelsep
    \usecounter{enumi}}
    \def\newblock{\hskip .11em plus .33em minus .07em}
    \sloppy\clubpenalty4000\widowpenalty4000
    \sfcode`\.=1000\relax}

\def\op#1{\mathop{\fam0 #1}\limits}

\newcommand{\beq}{\begin{equation}}
\newcommand{\eeq}{\end{equation}}
\newcommand{\ben}{\begin{eqnarray}}
\newcommand{\een}{\end{eqnarray}}
\newcommand{\be}{\begin{eqnarray*}}
\newcommand{\ee}{\end{eqnarray*}}
\newcommand{\bea}{\begin{eqalph}}
\newcommand{\eea}{\end{eqalph}}

\newcommand{\cT}{{\cal T}}

\newcommand{\al}{\alpha}
\newcommand{\vr}{\varrho}
\newcommand{\bt}{\beta}

\newcommand{\la}{\lambda}

\newcommand{\m}{\mu}

\newcommand{\g}{\gamma}
\newcommand{\G}{\Gamma}

\newcommand{\di}{{\rm dim\,}}

\newcommand{\si}{\sigma}
\newcommand{\Si}{\Sigma}
\newcommand{\w}{\wedge}
\newcommand{\wt}{\widetilde}

\newcommand{\ol}{\overline}
\newcommand{\dr}{\partial}
\newcommand{\ar}{\op\longrightarrow}
\newcommand{\ot}{\otimes}

\newcounter{theorem}
\newcounter{remark}
\newcounter{proposition}
\newcounter{lemma}
\newcounter{corollary}
\newcounter{definition}

\def\theremark{\arabic{remark}}

\def\thedefinition{\arabic{definition}}

\newenvironment{proof}{\noindent
{\bf Proof.}}{\hfill $\Box$ \medskip}
\newenvironment{rem}{\refstepcounter{remark}\medskip\noindent{\bf Remark
\theremark.}}{\medskip}

\newenvironment{theo}{\refstepcounter{definition} \medskip
\noindent{\bf Theorem \thedefinition.}}{\medskip}
\newenvironment{prop}{\refstepcounter{definition} \medskip
\noindent{\bf Proposition \thedefinition.}}{\medskip}

\newcommand{\mar}[1]{}

\begin{document}
\hbox{}

\begin{center}

{\large\bf GEOMETRY OF CLASSICAL HIGGS FIELDS}
\bigskip

{\sc G.SARDANASHVILY}

{\it Department of Theoretical Physics, Moscow State University,
117234, Moscow, Russia \\ sard@grav.phys.msu.su}

\end{center}

\begin{small}

\bigskip
In gauge theory, Higgs fields are responsible for spontaneous
symmetry breaking. In classical gauge theory on a principal bundle
$P$, a symmetry breaking is defined as the reduction of a
structure group of this principal bundle to a subgroup $H$ of
exact symmetries. This reduction takes place iff there exists a
global section of the quotient bundle $P/H$. It is a classical
Higgs field. A metric gravitational field exemplifies such a Higgs
field. We summarize the basic facts on the reduction in principal
bundles and geometry of Higgs fields. Our goal is the particular
covariant differential in the presence of a Higgs field.
\end{small}

\section{Introduction}

Gauge theory deals with the three types of classical fields. These
are gauge potentials, matter fields and Higgs fields. Higgs fields
are responsible for spontaneous symmetry breaking. Spontaneous
symmetry breaking is a quantum phenomenon. In classical gauge
theory on a principal bundle $P\to X$, a symmetry breaking is
defined as the reduction of the structure Lie group $G$, $\di
G>0$, of this principal bundle to a closed (consequently, Lie)
subgroup $H$, $\di H>0$, of exact symmetries
\cite{iva,tra,nik,perc,keyl,sard92}. From the mathematical
viewpoint, one speaks on the Klein--Chern geometry or a reduced
$G$-structure \cite{zul,kob72,gor}. By virtue of the well-known
theorem (see Theorem \ref{redsub} below), the reduction of the
structure group of a principal bundle takes place iff there exists
a global section of the quotient bundle $P/H\to X$. It is treated
as a classical Higgs field. In the gauge gravitation theory, a
pseudo-Riemannian metric exemplifies such a Higgs field,
associated to the Lorentz reduced structure \cite{iva,sardz92}.

This article aims to summarize the relevant material on the
reduction in principal bundles and geometry of Higgs fields. It is
geometry on the composite fiber bundle
\mar{b3223a}\beq
P\to P/H\to X, \label{b3223a}
\eeq
where
\mar{b3194}\beq
 P_\Si=P\ar^{\pi_{P\Si}} P/H \label{b3194}
\eeq
is a principal bundle with the structure group $H$ and
\mar{b3193}\beq
\Si=P/H\ar^{\pi_{\Si X}} X \label{b3193}
\eeq
is a $P$-associated fiber bundle with the typical fiber $G/H$ on
which the structure group $G$ acts on the left. Let $Y\to \Si$ be
a vector bundle associated to the $H$-principal bundle $P_\Si$
(\ref{b3194}). Then sections of the fiber bundle $Y\to X$ describe
matter fields with the exact symmetry group $H$ in the presence of
Higgs fields \cite{sard92}. A problem is that $Y\to X$ fails to be
a $P$-associated bundle with a structure group $G$ and,
consequently, it need not admit a principal connection. Our goal
is the particular covariant differential (\ref{b3260}) on $Y\to X$
defined by a principal connection on the $H$-principal bundle
$P\to P/H$, but not $P\to X$. For instance, this is the case of
the covariant differential of spinor fields in the gauge
gravitation theory \cite{sard98a,sard02}.

\section{Reduced structures}

We start with a few Remarks summarizing the relevant facts on
principal and associated bundles \cite{book00}.

\begin{rem} \label{mos60} \mar{mos60}
By a fiber bundle associated to the principal bundle $P\to X$ is
usually meant the quotient
\mar{b1.230}\beq
Y=(P\times V)/G, \label{b1.230}
\eeq
where the structure group $G$ acts on the typical fiber $V$ of $Y$
on the left. The quotient (\ref{b1.230}) is defined by
identification of the elements $(p,v)$ and $(pg,g^{-1}v)$ for all
$g\in G$. By $[p]$ is further denoted the restriction of the
canonical morphism
\be
 P\times V\to (P\times V)/G
\ee
to $\{p\}\times V$, and we write $[p](v)= (p,v)\cdot G$, $v\in V$.
Then, by definition of $Y$, we have $[p](v)=[pg](g^{-1}v)$.
Strictly speaking, $Y$ (\ref{b1.230}) is a fiber bundle
canonically associated to a principal bundle $P$. Recall that a
fiber bundle $Y\to X$, given by the triple $(X,V,\Psi)$ of a base
$X$, a typical fiber $V$ and a bundle atlas $\Psi$, is called a
fiber bundle with a structure group $G$ if $G$ acts effectively on
$V$ on the left and the transition functions $\rho_{\la\bt}$ of
the atlas $\Psi$ take their values into the group $G$. The set of
these transition functions form a cocycle. Atlases of $Y$ are
equivalent iff cocycles of their transition functions are
equivalent. The set of equivalent cocycles are elements of the
first cohomology set $H^1(X;G_\infty)$. Fiber bundles
$(X,V,G,\Psi)$ and $(X,V',G,\Psi')$ with the same structure group
$G$, which may have different typical fibers, are called
associated if the transition functions of the atlases $\Psi$ and
$\Psi'$ belong to the same element of the the cohomology set
$H^1(X;G_\infty)$. Any two associated fiber bundles with the same
typical fiber are isomorphic to each other, but their isomorphism
is not canonical in general. A fiber bundle $Y\to X$ with a
structure group $G$ is associated to some $G$-principal bundle
$P\to X$. If $Y$ is canonically associated to $P$ as in
(\ref{b1.230}), then every atlas of $P$ determines canonically the
associated atlas of $Y$, and every automorphism of a principal
bundle $P$ yields the corresponding automorphism of the
$P$-associated fiber bundle (\ref{b1.230}).
\end{rem}

\begin{rem}
Recall that an automorphism $\Phi_P$ of a principal bundle $P$, by
definition, is equivariant under the canonical action
$R_g\circ\Phi_P=\Phi_P\circ R_g$, $g\in G$, of the structure group
$G$ on $P$. Every automorphism of $P$ yields the corresponding
automorphisms
\mar{024}\beq
\Phi_Y: (p,v)\cdot G\mapsto  (\Phi_P(p),v)\cdot G, \qquad p\in P,
\qquad v\in V, \label{024}
\eeq
of the $P$-associated bundle $Y$ (\ref{b1.230}). For the sake of
brevity, we will write
\be
\Phi_Y: (P\times V)/G\to (\Phi_P(P)\times V)/G.
\ee
Every automorphism of a principal bundle $P$ is represented as
\mar{b3111}\beq
\Phi_P(p)=pf(p), \qquad p\in P, \label{b3111}
\eeq
where $f$ is a $G$-valued equivariant function on $P$, i.e.,
$f(pg)=g^{-1}f(p)g$, $g\in G$.
\end{rem}

One says that the structure group $G$ of a principal bundle $P$ is
reducible to a Lie subgroup $H$ if there exists a $H$-principal
subbundle $P^h$ of $P$ with the structure group $H$. This
subbundle is called a reduced $G^\downarrow H$-structure. Two
reduced $G^\downarrow H$-structures $P^h$ and $P^{h'}$ on a
$G$-principal bundle are said to be isomorphic if there is an
automorphism $\Phi$ of $P$ which provides an isomorphism of $P^h$
and $P^{h'}$. If $\Phi$ is a vertical automorphism of $P$, reduced
structures $P^h$ and $P^{h'}$ are called equivalent.

\begin{rem}
Note that, in \cite{kob72,gor} (see also \cite{cord}), the reduced
structures on the principle bundle $LX$ of linear frames in the
tangent bundle $TX$ of $X$ are only considered, and a class of
isomorphisms of such reduced structures is restricted to holonomic
automorphisms of $LX$, i.e., the canonical lifts onto $LX$ of
diffeomorphisms of the base $X$.
\end{rem}

Let us recall the following two theorems \cite{kob}.

\begin{theo}
A structure group $G$ of a principal bundle $P$ is reducible to
its closed subgroup $H$ iff $P$ has an atlas $\Psi_P$ with
$H$-valued transition functions.
\end{theo}

Given a reduced subbundle $P^h$ of $P$, such an atlas $\Psi_P$ is
defined by a family of local sections $\{z_\al\}$ which take their
values into $P^h$.

\begin{theo}\label{redsub} \mar{redsub}
There is one-to-one correspondence $P^h=\pi_{P\Si}^{-1}(h(X))$
between the reduced $H$-principal subbundles $P^h$ of $P$ and the
global sections $h$ of the quotient fiber bundle $P/H\to X$.
\end{theo}

In general, there are topological obstructions to the reduction of
a structure group of a principal bundle to its subgroup. One
usually refers to the following theorems \cite{ste}.

\begin{theo} \label{mos9} \mar{mos9}
Any fiber bundle $Y\to X$ whose typical fiber is
diffeomorphic to $\Bbb R^m$ has a global section. Its section over
a closed subset $N\subset X$ is always extended to a global
section.
\end{theo}

By virtue of Theorem \ref{mos9}, the structure group $G$ of a
principal bundle $P$ is always reducible to its closed subgroup
$H$ if the quotient $G/H$ is diffeomorphic to a Euclidean space.

\begin{theo}
A structure group $G$ of a principal bundle is always reducible to
its maximal compact subgroup $H$ since the quotient space $G/H$ is
homeomorphic to a Euclidean space.
\end{theo}

For instance, this is the case of $G=GL(n,\Bbb C)$, $H=U(n)$ and
$G=GL(n,\Bbb R)$, $H=O(n)$. In the last case, the associated Higgs
field is a Riemannian metric on $X$.

It should be emphasized that different $H$-principal subbundles
$P^h$ and $P^{h'}$ of a $G$-principal bundle $P$ need not be
isomorphic to each other in general.

\begin{prop}
Every vertical automorphism $\Phi$ of a principal bundle $P\to X$
sends an $H$-principal subbundle $P^h$ onto an equivalent
$H$-principal subbundle $P^{h'}$. Conversely, let two reduced
subbundles $P^h$ and $P^{h'}$ of a principal bundle $P$ be
isomorphic to each other, and $\Phi:P^h\to P^{h'}$ be an
isomorphism over $X$. Then $\Phi$ is extended to a vertical
automorphism of $P$.
\end{prop}

\begin{proof}
 Let
\be
\Psi^h=\{(U_\al,z^h_\al), \rho^h_{\al\bt}\}, \qquad z^h_\al(x) =
z^h_\bt(x)\rho^h_{\al\bt}(x), \qquad x\in U_\al\cap U_\bt,
\ee
be an atlas of the reduced subbundle $P^h$, where $z^h_\al$ are
local sections of $P^h\to X$  and $\rho^h_{\al\bt}$ are the
transition functions. Given a vertical automorphism $\Phi$ of $P$,
let us provide the reduced subbundle $P^{h'}=\Phi(P^h)$ with the
atlas $\Psi^{h'}=\{(U_\al,z^{h'}_\al), \rho^{h'}_{\al\bt}\}$
determined by the local sections $z^{h'}_\al =\Phi\circ z^h_\al$
of $P^{h'}\to X$. Then it is readily observed that
$\rho^{h'}_{\al\bt}(x) =\rho^h_{\al\bt}(x)$, $x\in U_\al\cap
U_\bt$. Conversely, any isomorphism $\Phi$ of reduced structures
$P^h$ and $P^{h'}$ on $P$ determines a $G$-valued  function $f$ on
$P^h$ given by the relation $pf(p)=\Phi(p)$, $p\in P^h$.
Obviously, this function is $H$-equivariant. Its prolongation to a
$G$-equivariant function on $P$ is defined to be
$f(pg)=g^{-1}f(p)g$, $p\in P^h$, $g\in G$. In accordance with the
relation (\ref{b3111}), this function defines a vertical
automorphism of $P$ whose restriction to $P^h$ coincides with
$\Phi$.
\end{proof}

\begin{prop} If the quotient $G/H$ is diffeomorphic to a
Euclidean space $\Bbb R^k$, all $H$-principal subbundles of a
$G$-principal bundle $P$ are equivalent to each other \cite{ste}.
\end{prop}

Given a reduced subbundle $P^h$ of a principal bundle $P$, let
\mar{b3.3000}\beq
Y^h=(P^h\times V)/H \label{b3.3000}
\eeq
be the associated fiber bundle with a typical fiber $V$. Let
$P^{h'}$ be another reduced subbundle of $P$ which is isomorphic
to $P^h$, and
\be
Y^{h'}=(P^{h'}\times V)/H
\ee
The fiber bundles $Y^h$ and $Y^{h'}$ are isomorphic, but not
canonically isomorphic in general.

\begin{prop}\label{iso2} \mar{iso2}  Let $P^h$ be a
$H$-principal subbundle of a $G$-principal bundle $P$. Let $Y^h$
be the $P^h$-associated bundle (\ref{b3.3000}) with a typical
fiber $V$. If $V$ carries a representation of the whole group $G$,
the fiber bundle $Y^h$ is canonically isomorphic to the
$P$-associated fiber bundle
\be
Y=(P\times V)/G.
\ee
\end{prop}

\begin{proof} Every element of $Y$ can be represented as $(p,v)\cdot
G$, $p\in P^h$. Then the desired isomorphism is
\be
Y^h\ni (p,v)\cdot H\quad \Longleftrightarrow \quad (p,v)\cdot G\in
Y.
\ee
It follows that, given a $H$-principal subbundle $P^h$ of $P$, any
$P$-associated fiber bundle $Y$ with the structure group $G$ is
canonically equipped with a structure of the $P^h$-associated
fiber bundle $Y^h$ with the structure group $H$. Briefly, we can
write
\be
Y=(P\times V)/G \simeq(P^h\times V)/H=Y^h.
\ee
However, if $P^h\neq P^{h'}$, the $P^h$- and $P^{h'}$-associated
bundle structures on $Y$ need not be equivalent. Given  bundle
atlases $\Psi^h$ of $P^h$ and $\Psi^{h'}$ of $P^{h'}$, the union
of the associated atlases of $Y$ has necessarily $G$-valued
transition functions between the charts of $\Psi^h$ and
$\Psi^{h'}$.
\end{proof}

\section{Classical Higgs fields}

In accordance with Theorem \ref{redsub}, the set of reduced
$H$-principal subbundles $P^h$ of $P$ is in bijective
correspondence with the set of Higgs fields $h$. Given such a
subbundle $P^h$, let $Y^h$ (\ref{b3.3000}) be the associated
vector bundle with a typical fiber $V$ which admits a
representation of the group $H$ of exact symmetries, but not the
whole symmetry group $G$. Its sections $s_h$ describe matter
fields in the presence of the Higgs fields $h$ and some principal
connection $A_h$ on $P^h$. In general, the fiber bundle $Y^h$
(\ref{b3.3000}) is not associated or canonically associated (see
Remark \ref{mos60}) to other $H$-principal subbundles $P^{h'}$ of
$P$. It follows that, in this case, $V$-valued matter fields can
be represented only by pairs with
 Higgs fields. The goal is to describe the totality of these pairs
$(s_h,h)$ for all Higgs fields $h$. We refer to the following
theorems \cite{book00}.

\begin{theo}\label{comp10} \mar{comp10}
Given an arbotrary composite fiber bundle
\mar{1.34}\beq
Y\ar^{\pi_{Y\Si}} \Si\ar^{\pi_{\Si X}} X, \label{1.34}
\eeq
let $h$ be a global section of the fiber bundle $\Si\to X$. Then
the restriction
\mar{S10}\beq
Y_h=h^*Y \label{S10}
\eeq
of the fiber bundle $Y\to\Si$ to $h(X)\subset \Si$ is a subbundle
$i_h: Y_h\hookrightarrow Y$ of the fiber bundle $Y\to X$.
\end{theo}

In the case of a principal bundle $Y=P$ and $\Si=P/H$, the
restriction $h^*P_\Si$ (\ref{S10}) of the $H$-principal bundle
$P_\Si$ (\ref{b3194}) to $h(X)\subset \Si$ is a $H$-principal
bundle over $X$, which is equivalent to the reduced subbundle
$P^h$ of $P$.

\begin{theo} \label{mos61} \mar{mos61}  Given
a section $h$ of the fiber bundle $\Si\to X$ and a section $s_\Si$
of the fiber bundle $Y\to\Si$, their composition
\mar{1.37}\beq
s=s_\Si\circ h \label{1.37}
\eeq
is a section of the composite fiber  bundle $Y\to X$ (\ref{1.34}).
Conversely, every section $s$ of the fiber bundle $Y\to X$ is the
composition (\ref{1.37}) of the section $h=\pi_{Y\Si}\circ s$ of
the fiber bundle $\Si\to X$ and some section $s_\Si$ of the fiber
bundle $Y\to \Si$ over the closed submanifold $h(X)\subset \Si$.
\end{theo}

Let us consider the composite fiber bundle (\ref{b3223a}) and the
composite fiber bundle
\mar{b3225}\beq
Y\ar^{\pi_{Y\Si}} P/H\ar^{\pi_{\Si X}} X \label{b3225}
\eeq
where $Y\to \Si=P/H$ is a vector bundle
\be
Y=(P\times V)/H
\ee
associated to the corresponding $H$-principal bundle $P_\Si$
(\ref{b3194}). Given a global section $h$ of the fiber bundle
$\Si\to X$ (\ref{b3193}) and the $P^h$-associated fiber bundle
(\ref{b3.3000}), there is the canonical injection
\be
i_h: Y^h=(P^h\times V)/H \hookrightarrow Y
\ee
over $X$ whose image is the restriction
\be
h^*Y=(h^*P\times V)/H
\ee
of the fiber bundle $Y\to\Si$ to $h(X)\subset \Si$, i.e.,
\mar{b3226}\beq
i_h(Y^h)\cong \pi^{-1}_{Y\Si}(h(X)) \label{b3226}
\eeq
(see Theorem \ref{comp10}). Then, by virtue of Theorem
\ref{mos61}, every global section $s_h$ of the fiber bundle $Y^h$
corresponds to the global section $i_h\circ s_h$ of the composite
fiber bundle (\ref{b3225}). Conversely, every global section $s$
of the composite fiber bundle (\ref{b3225}) which projects onto a
section $h=\pi_{Y\Si}\circ s$ of the fiber bundle $P/H\to X$ takes
its values into the subbundle $i_h(Y^h)\subset Y$ in accordance
with the relation (\ref{b3226}). Hence, there is one-to-one
correspondence between the sections of the fiber bundle $Y^h$
(\ref{b3.3000}) and the sections of the composite fiber bundle
(\ref{b3225}) which cover $h$.

Thus, it is precisely the composite fiber bundle (\ref{b3225})
whose sections describe the above-mentioned totality of pairs
$(s_h, h)$ of matter fields and Higgs fields in gauge theory with
broken symmetries. For instance, this is the case of spinor fields
in the presence of gravitational fields \cite{sard98a}. A problem
is that the typical fiber of the fiber bundle $Y\to X$ fails to
admit a representation of the group $G$, unless $G\to G/H$ is a
trivial bundle. It follows that $Y\to X$ is not associated to $P$
and, it does not admit a principal connection in general. If $G\to
G/H$ is a trivial bundle, there exists its global section whose
values are representatives of elements of $G/H$. In this case, the
typical fiber of $Y\to X$ is $V\times G/H$, and one can provide it
with an induced representation of $G$. Of course, this
representation is not canonical, unless $V$ itself admits a
representation of $G$.

\section{Composite and reduced connections}

Since the reduction in a principal bundle leads to the composite
fiber bundle (\ref{b3223a}), we turn to the notion of a composite
connection \cite{book00}.

Let us consider the composite bundle (\ref{1.34}) provided with
bundle coordinates $(x^\la,\si^m,y^i)$, where $(x^\m,\si^m)$ are
bundle coordinates on $\Si\to X$ and the transition functions of
$\si^m$ are independent of the coordinates $y^i$. Let us consider
the jet manifolds $J^1\Si$, $J^1_\Si Y$, and $J^1Y$ of the fiber
bundles $\Si\to X$, $Y\to \Si$ and $Y\to X$, respectively. They
are endowed with the adapted coordinates
\be
( x^\la ,\si^m, \si^m_\la),\quad ( x^\la ,\si^m, y^i, \wt y^i_\la,
y^i_m),\quad ( x^\la ,\si^m, y^i, \si^m_\la ,y^i_\la).
\ee
There is the canonical map \cite{sau}
\be
\vr : J^1\Si\op\times_\Si J^1_\Si Y\ar_Y J^1Y, \qquad
y^i_\la\circ\vr=y^i_m{\si}^m_{\la} +\wt y^i_{\la}.
\ee

With this map, one can obtain the relations between connections on
the fiber bundles $Y\to X$, $Y\to\Si$ and $\Si\to X$ as follows.
Let
\mar{b1.113,1}\ben
&& A_\Si=dx^\la\ot (\dr_\la + A_\la^i\dr_i) +d\si^m\ot (\dr_m +
A_m^i\dr_i), \label{b1.113}\\
&& \G=dx^\la\ot (\dr_\la + \G_\la^m\dr_m) \label{b1.111}
\een
be connections on the fiber bundles $Y\to \Si$ and $\Si\to X$,
respectively. They define the composite connection
\mar{b1.114}\beq
\g=dx^\la\ot (\dr_\la +\G_\la^m\dr_m + (A_\la^i +
A_m^i\G_\la^m)\dr_i) \label{b1.114}
\eeq
on $Y\to X$ in accordance with the diagram
\be
\begin{array}{rcccl}
 & J^1\Si\op\times_\Si J^1Y_\Si & \ar^\vr & J^1Y & \\
_{(\G,A_\Si)} & \put(0,-10){\vector(0,1){20}} & &
\put(0,-10){\vector(0,1){20}} & _{\g} \\
 & \Si\op\times_X Y & \longleftarrow & Y &
\end{array}
\ee
In brief, we will write
\mar{b1.500}\beq
\g=A_\Si\circ\G. \label{b1.500}
\eeq
In particular, let us consider a vector field $\tau$ on the base
$X$, its horizontal lift $\G\tau$ onto $\Si$ by means of the
connection $\G$ and, in turn, the horizontal lift $A_\Si(\G\tau)$
of $\G\tau$ onto $Y$ by means of the connection $A_\Si$. Then
$A_\Si(\G\tau)$ coincides with the horizontal lift $\g\tau$ of
$\tau$ onto $Y$ by means of the composite connection $\g$
(\ref{b1.500}).

\begin{rem} Recall the notions of a pull-back connection and
a reducible connection. Given a fiber bundle $Y\to X$, let
$f:X'\to X$ be a map and $f^*Y\to X'$ the pull-back of $Y$ by $f$.
Written as a vertical-valued form
\be
 \G= (dy^i -\G^i_\la dx^\la)\ot\dr_i,
\ee
any connection $\G$ on $Y\to X$ yields the pull-back connection
\mar{mos82}\beq
f^*\G=(dy^i-(\G\circ f_Y)^i_\la\frac{\dr f^\la}{\dr
x'^\m}dx'^\m)\ot\dr_i \label{mos82}
\eeq
on $f^*Y\to X'$. In particular, let $P$ be a principal bundle and
$f^*P$ the pull-back principal bundle with the same structure
group. If $A$ is a principal connection on $P$, then the pull-back
connection $f^*A$ (\ref{mos82}) on $f^*P$ is also a principal
connection \cite{kob}. Let $i_Y:Y\to Y'$ be a subbundle of a fiber
bundle  $Y'\to X$ and $\G'$ a connection on $Y'\to X$. If there
exists a connection $\G$ on $Y\to X$ such that the diagram
\be
\begin{array}{rcccl}
 & Y' & \op\longrightarrow^{\G'} & J^1Y &  \\
 _{i_Y} & \put(0,-10){\vector(0,1){20}} & & \put(0,-10){\vector(0,1){20}} &
_{J^1i_Y} \\
 & Y & \op\longrightarrow^\G & {J^1Y'} &
 \end{array}
 \ee
commutes, we say that $\G'$ is {\sl reducible}  to the connection
$\G$.\index{reducible connection} The following conditions are
equivalent:

(i) $\G'$ is reducible to $\G$;

(ii) $Ti_Y(HY)=HY'\vert_{i_Y(Y)}$, where $HY\subset TY$ and
$HY'\subset TY'$ are the horizontal subbundles determined by $\G$
and $\G'$, respectively;

(iii) for every vector field $\tau\in\cT(X)$, the vector fields
$\G\tau$ and $\G'\tau$ are $i_Y$-related, \index{related vector
fields} i.e.,
\mar{b1.117}\beq
Ti_Y\circ\G \tau=\G' \tau\circ i_Y.  \label{b1.117}
\eeq
\end{rem}

Let $h$ be a section of the fiber bundle $\Si\to X$ and $Y_h$ the
subbundle (\ref{S10}) of the composite fiber bundle $Y\to X$,
which is the restriction of the fiber bundle $Y\to\Si$ to $h(X)$.
Every connection $A_\Si$ (\ref{b1.113}) induces the pull-back
connection
\mar{mos83}\beq
A_h=i_h^*A_\Si=dx^\la\ot[\dr_\la+((A^i_m\circ h)\dr_\la h^m
+(A\circ h)^i_\la)\dr_i] \label{mos83}
\eeq
on $Y_h\to X$. Now, let $\G$ be a connection on $\Si\to X$ and let
$\g=A_\Si\circ\G$ be the composition (\ref{b1.500}). Then  it
follows  from (\ref{b1.117}) that the connection $\g$ is reducible
to the connection $A_h$ iff the section $h$ is an integral section
of $\G$, i.e., $\G_\la^m\circ h=\dr_\la h^m$. Such a connection
$\G$ always exists.

Given a composite fiber bundle $Y$ (\ref{1.34}), there is the
following exact sequences
\mar{63}\beq
0\to V_\Si Y\hookrightarrow VY\to Y\op\times_\Si V\Si\to 0,
\label{63a}
\eeq
of vector bundles over $Y$, where $V_\Si Y$ denotes the vertical
tangent bundle of the fiber bundle $Y\to\Si$. Every connection $A$
(\ref{b1.113}) on the fiber bundle $Y\to\Si$ determines the
splitting
\mar{46a}\ben
&& VY=V_\Si Y\op\oplus_Y A_\Si(Y\op\times_\Si V\Si),\label{46a}\\
&& \dot y^i\dr_i + \dot\si^m\dr_m= (\dot y^i -A^i_m\dot\si^m)\dr_i
+ \dot\si^m(\dr_m+A^i_m\dr_i), \nonumber
\een
of the exact sequence (\ref{63a}). Using this splitting, one can
construct the first order differential operator
\mar{7.10}\beq
D: J^1Y\to T^*X\op\otimes_Y V_\Si Y, \qquad D=
dx^\la\otimes(y^i_\la- A^i_\la -A^i_m\si^m_\la)\dr_i, \label{7.10}
\eeq
called the vertical covariant differential on the composite fiber
bundle $Y\to X$. It possesses the following important property.
Let $h$ be a section of the fiber bundle $\Si\to X$ and $Y_h$ the
subbundle (\ref{S10}) of the composite fiber bundle $Y\to X$,
which is the restriction of the fiber bundle $Y\to\Si$ to $h(X)$.
Then the restriction of the vertical covariant differential $D$
(\ref{7.10}) to $J^1i_h(J^1Y_h)\subset J^1Y$ coincides with the
familiar covariant differential on $Y_h$ relative to the pull-back
connection $A_h$ (\ref{mos83}).

Turn now to the properties of connections compatible with a
reduced structure of a principal bundle. Recall the following
theorems \cite{kob}.

\begin{theo} \label{mos176} \mar{mos176}
Since principal connections are equivariant, every principal
connection $A_h$ on a reduced $H$-principal subbundle $P^h$ of a
$G$-principal bundle $P$ gives rise to a principal connection on
$P$.
\end{theo}

\begin{theo}
A principal connection $A$ on a $G$-principal bundle $P$ is
reducible to a principal connection on a reduced $H$-principal
subbundle $P^h$ of $P$ iff the corresponding global section $h$ of
the $P$-associated fiber bundle $P/H\to X$ is an integral section
of the associated principal connection $A$ on $P/H\to X$.
\end{theo}

\begin{theo}
Let the Lie algebra ${\got g}(G)$ of $G$ is the direct sum
\be
{\got g}(G) = {\got g}(H) \oplus {\got m}
\ee
of the Lie algebra ${\got g}(H)$ of $H$ and a subspace ${\got m}$
such that $ad(g)({\got m})\subset {\got m}$, $g\in H$. Let $\ol A$
be a ${\got g}(G)$-valued connection form on $P$. Then, the
pull-back of the ${\got g}(H)$-valued component of $\ol A$ onto a
reduced subbundle $P^h$ is the connection form of a principal
connection on $P^h$.
\end{theo}

Given the composite fiber bundle (\ref{b3223a}), let $A_\Si$ be a
principal connection on the $H$-principal bundle $P\to P/H$. Then,
for any reduced $H$-principal subbundle $P^h$ of $P$, the
pull-back connection $i_h^*A_\Si$ (\ref{mos83}) is a principal
connection on $P^h$. With this fact, we come to the following
feature of the dynamics of field systems with symmetry breaking.

Let the composite fiber bundle $Y$ (\ref{b3225}) be provided with
coordinates $(x^\la, \si^m, y^i)$, where $(x^\la, \si^m)$ are
bundle coordinates on the fiber bundle $\Si\to X$. As was
mentioned above, the fiber bundle $Y\to X$ fails to be a
$P$-associated bundle in general and, consequently, need not admit
a principal connection. Therefore, let us consider a principal
connection $A_\Si$ (\ref{b1.113}) on the vector bundle $Y\to \Si$.
This connection defines the splitting (\ref{46a}) of the vertical
tangent bundle $VY$ and leads to the vertical covariant
differential $D$ (\ref{7.10}). As was mentioned above, this
operator possesses the following property. Given a global section
$h$ of $\Si\to X$, its restriction
\mar{b3260}\ben
&& D_h =D\circ J^1i_h: J^1Y^h \to T^*X\ot VY^h, \label{b3260}\\
&& D_h =dx^\la\ot(y^i_\la- A^i_\la -A^i_m\dr_\la h^m)\dr_i,
\nonumber
\een
to $Y^h$ is precisely the familiar covariant differential relative
to the pull-back principal connection $A_h$ (\ref{mos83}) on the
fiber bundle $Y^h\to X$. As a consequence, one can construct a
Lagrangian on the jet manifold $J^1Y$ of a composite fiber bundle
$Y$ (\ref{b3225}) which factorizes through the vertical covariant
differential $D$ as
\be
L:J^1Y\op\to^{D}T^*X\op\ot_YVY_\Si\to\op\w^nT^*X.
\ee

\begin{rem}
Given a $G$-principal bundle $P$, let $P^h$ be its reduced
$H$-principal subbundle. Let $A_\Si$ be a principal connection on
the $H$-principal bundle $P\to P/H$, and $i_h^*A_\Si$
(\ref{mos83}) the pull-back principal connection on $P^h$. In
accordance with Theorem \ref{mos176}, it gives rise to a principal
connection on $P$. For different $h$ and $h'$, the connections
$i_h^*A_\Si$ and $i_{h'}^*A_\Si$ however yield different principal
connections on $P$.
\end{rem}

\end{document}